\documentclass{article}
\usepackage{spconf,amsmath,graphicx}
\usepackage{epsfig,graphicx,subfigure,psfrag,amsmath,cases}
\usepackage{latexsym,amssymb,amsmath,epsfig,subfigure,algorithm,mathtools}
\usepackage{algorithmic}
\usepackage{color}
\usepackage{url}
\usepackage{scrtime}
\usepackage{multicol}

\newtheorem{T-Prob}{Transformed Problem}

\DeclareMathOperator{\maxo}{maximize}
\DeclareMathOperator{\mino}{minimize}

\newcommand{\abs}[1]{\lvert#1\rvert}

\title{Joint Power and Subcarrier Allocation for Multicarrier Full-Duplex Systems}

\name{Yan Sun$^1$, Derrick Wing Kwan Ng$^2$, and Robert Schober$^1$\thanks{Robert Schober is also with the University of British Columbia, Vancouver,  Canada.  }}
\address{$^1$Institute for Digital Communications \\ Friedrich-Alexander-University Erlangen-N\"urnberg (FAU), Germany\\
$^2$School of Electrical Engineering and Telecommunications \\ The University
of New South Wales, Australia\vspace*{-3mm}}

\begin{document}
\maketitle
\begin{abstract}
In this paper, we investigate resource allocation for multicarrier communication systems employing a full-duplex base station for serving multiple half-duplex downlink and uplink  users simultaneously. We study the joint power and subcarrier allocation design for the maximization of the weighted sum throughput of the system. The algorithm design is formulated as a mixed combinatorial non-convex optimization problem and obtaining the globally optimal solution may require prohibitively
high computational complexity. Therefore, a low computational complexity suboptimal iterative algorithm exploiting  successive convex approximation is proposed to obtain a locally optimal solution. Simulation results confirm that the proposed suboptimal algorithm obtains a substantial improvement in system throughput compared to various existing baseline schemes.
\end{abstract}
%
\renewcommand{\baselinestretch}{0.95}
\large\normalsize
\vspace*{-5mm}
\section{Introduction}
\label{sec:intro}
Multicarrier (MC) communication techniques have been widely investigated over the last decades, since they enable flexible resource allocation, e.g. dynamic subcarrier and power allocation, for serving multiple users \cite{book:david_wirelss_com}--\nocite{JR:Linglong_OFDM,JR:Linglong_OFDM2,JR:Chen_MC,JR:Roger_OFDMA,JR:QQ_OFDM_EE,ng2012energy,sun2016optimal, CN:MC_NOMA_Zhiqiang}\cite{CN:MC_NOMA_Zhiqiang}.
In \cite{ng2012energy}, the authors studied the resource allocation algorithm design for energy-efficient communication in multi-cell orthogonal frequency division multiple access systems.
In \cite{sun2016optimal} and \cite{CN:MC_NOMA_Zhiqiang}, the joint power and subcarrier allocation algorithm design was investigated for weighted sum throughput maximization and power consumption minimization in MC non-orthogonal
multiple access (NOMA) systems, respectively.
However, despite the fruitful development of MC-based communication in \cite{book:david_wirelss_com}--\nocite{JR:QQ_OFDM_EE,ng2012energy,sun2016optimal, CN:MC_NOMA_Zhiqiang}\cite{CN:MC_NOMA_Zhiqiang},  radio  resources are  not efficiently utilized, since  base stations (BSs) operate in the half-duplex (HD) mode, where uplink (UL) and downlink (DL) transmissions are separated by orthogonal radio resources in either time or frequency leading to resource underutilization.

Recently, full-duplex (FD) communication has become an emerging technique for the fifth-generation (5G) communication networks which is capable of potentially doubling the spectral efficiency by performing simultaneous DL and UL transmission in the same frequency band \cite{Sun16FDSecurity,Kwan2016FDDAS}.
Therefore, it is expected that the spectral efficiency of conventional HD-MC systems can be improved substantially by employing an FD BS.
However,  self-interference (SI) at the FD BS and co-channel interference (CCI) between DL and UL users on each subcarrier may degrade the quality of service (QoS) in FD communication systems. Therefore, different resource allocation designs for FD MC systems were proposed to overcome these challenges \cite{Dynamic_FDrelay,Nam15Joint}.
In \cite{Dynamic_FDrelay}, the authors proposed an optimal joint precoding and scheduling algorithm for the maximization of the weighted sum throughput of MIMO-MC-FD relaying systems. However, the optimal solution in \cite{Dynamic_FDrelay} can only be obtained when the number of subcarriers approaches infinity which may not be achievable in practice.
The authors of \cite{Nam15Joint} proposed a suboptimal iterative subcarrier and power allocation algorithm for the maximization of the weighted sum throughput in an FD MC system.
However, the iterative approach proposed in  \cite{Nam15Joint} divided the optimization problem into individual uplink and downlink subproblems which may lead to degradation in performance.

Motivated by the aforementioned observations, we formulate a mixed combinatorial non-convex optimization problem to maximize the weighted system sum throughput of an FD MC system. In order to strike a balance between computational complexity and optimality, we propose a suboptimal joint subcarrier and power allocation algorithm based on successive convex approximation to obtain a locally optimal solution.

\vspace*{-5mm}
\section{System Model}
\label{sec:format}
In this section, we present the adopted notation and the considered FD MC system model.
\vspace*{-3mm}
\subsection{Notation}%
We use boldface lower case letters to denote vectors. $\mathbf{a}^T$ denotes the transpose of vector $\mathbf{a}$; $\mathbb{C}$ denotes the set of complex values; $\mathbb{R}$ denotes the set of non-negative real values; $\mathbb{R}^{N\times 1}$ denotes the set of all $N\times 1$ vectors with real entries and $\mathbb{R}^{N\times 1}_{\mathrm{+}}$ denotes the non-negative subset of $\mathbb{R}^{N\times 1}$; $\mathbb{Z}^{N\times 1}$ denotes the set of all $N\times 1$ vectors with integer entries; $\mathbf{a} \le \mathbf{b}$ indicates that $\mathbf{a}$ is component-wise smaller than $\mathbf{b}$; $\abs{\cdot}$ denotes the absolute value of a complex scalar; ${\cal E}\{\cdot\}$ denotes statistical expectation. The circularly symmetric complex Gaussian distribution with mean $w$ and variance $\sigma^2$ is denoted by ${\cal CN}(w,\sigma^2)$; and $\sim$ stands for ``distributed as". $\nabla_{\mathbf{x}} f(\mathbf{x})$ denotes the gradient vector of function $f(\mathbf{x})$ whose components are the partial derivatives of $f(\mathbf{x})$.

\vspace*{-3mm}
\subsection{FD MC System}
\begin{figure}[t]
 \centering
\includegraphics[width=2.5in]{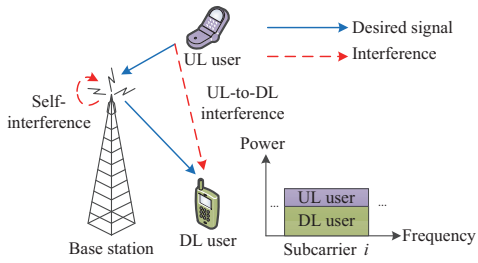} \vspace*{-2mm}
\caption{An FD MC system where an FD BS serves $K = 1$ HD DL user and $J = 1$ HD UL user on ${N_{\mathrm{F}}}$ subcarriers.}
\label{fig:FD-MC-model}\vspace*{-4mm}
\end{figure}\vspace*{-0mm}
We consider a FD MC system which consists of an FD BS, $K$ HD DL users, and $J$ HD UL users.  All transceivers are equipped with a single antenna. The entire frequency band of $W$ Hertz is partitioned into ${N_{\mathrm{F}}}$ orthogonal subcarriers. In this paper, we assume that each subcarrier is allocated to at most one DL user and one UL user on each subcarrier, cf. Figure \ref{fig:FD-MC-model}.
Assuming DL user $m\in \{1,\ldots,K\}$ and UL user $r\in \{1,\ldots,J\}$ are selected and multiplexed on subcarrier $i \in \{1,\ldots,N_{\mathrm{F}}\}$, the received signals at DL user $m$ and the FD BS are given by \vspace*{-2mm}
\begin{eqnarray}
\hspace*{-5mm} y_{\mathrm{DL}_m}^{i}\hspace*{-3.5mm}&=&\hspace*{-3.5mm}\sqrt{p_m^i \varpi_m}h_m^i x_{\mathrm{DL}_m}^{i}  + \hspace*{-3mm}
\underbrace{\sqrt{q_r^i \vartheta_{r,m}} f_{r,m}^i x_{\mathrm{UL}_r}^{i}}_{\mbox{co-channel interference}} \hspace*{-2mm} + z_{\mathrm{DL}_m}^{i},\\
\hspace*{-5mm} y_{\mathrm{BS}}^{i}\hspace*{-3.5mm}&=&\hspace*{-3.5mm}\sqrt{q_r^i \varrho_r}g_r^i x_{\mathrm{UL}_r}^{i}  + \underbrace{l_{\mathrm{SI}}^i \sqrt{p_m^i} x_{\mathrm{DL}_m}^{i} }_{\mbox{self-interference}} + \hspace*{1mm} z_{\mathrm{BS}}^{i},  \,\,\,\,
\end{eqnarray}
respectively. Variables $x_{\mathrm{DL}_m}^{i}\in\mathbb{C}$ and $x_{\mathrm{UL}_r}^{i}\in\mathbb{C}$ denote the symbols transmitted by the FD BS to DL user $m$ and by UL user $r$ to the FD BS on subcarrier $i$, respectively. Besides, without loss of generality, ${\cal E}\{\abs{x_{\mathrm{DL}_m}^{i}}^2\}={\cal E}\{\abs{x_{\mathrm{UL}_r}^{i}}^2\}=1, \forall m,r$ is assumed. $p_m^i$ is the transmit power of the signal intended for DL user $m$ at the FD BS and $q_r^i$ is the transmit power of the signal intended for the FD BS at UL user $r$ on subcarrier $i$. Variables $h_m^i\in\mathbb{C}$, $g_r^i\in\mathbb{C}$, and $f_{r,m}^i\in\mathbb{C}$ denote the small scale fading coefficients for the link between the FD BS and DL user $m$, the link between UL user $r$ and the FD BS, and the link between UL user $r$ and DL user $m$ on subcarrier $i$, respectively.
$l_{\mathrm{SI}}^i\in\mathbb{C}$ denotes the SI channel at the FD BS.
Variables $\varpi_m\in\mathbb{R}$, $\varrho_r\in\mathbb{R}$, and $\vartheta_{r,m}\in\mathbb{R}$ represent the joint effect of path loss and shadowing between the FD BS and DL user $m$, between UL user $r$ and the FD BS, and between UL user $r$ and DL user $m$, respectively. $z_{\mathrm{DL}_m}^i\sim{\cal CN}(0,\sigma_{\mathrm{z}_{\mathrm{DL}_m}}^2)$ and $z_{\mathrm{BS}}^i\sim{\cal CN}(0,\sigma_{\mathrm{z}_{\mathrm{BS}}}^2)$ denote the complex additive white Gaussian noise (AWGN) on subcarrier $i$ at DL user $m$ and the FD BS, respectively. Besides, for the study of resource allocation algorithm design, we assume that the global channel state information (CSI) of all links in the network is available at the BS so as to unveil the performance upper bound of practical FD MC systems.

\vspace*{-3mm}
\section{Problem Formulation and Solution}
In this section, after defining the adopted performance measure, we formulate the resource allocation problem. Then, we propose a iterative algorithm to solve the proposed problem.
\vspace*{-3mm}
\subsection{Weighted System Throughput}
Assuming DL user $m$ and UL user $r$ are allocated on subcarrier $i$, the weighted sum throughput on subcarrier $i$ is:
\begin{eqnarray} \label{throughput_i}
&&\hspace*{-5mm}U_{m,r}^i(p_{m}^i,q_{r}^i,s_{m,r}^{i}) \\
\hspace*{-0.5mm}=&&\hspace*{-5mm}\hspace*{-1.5mm} s_{m,r}^{i} \hspace*{-0.5mm} \Bigg[\hspace*{-0.5mm}
w_m \hspace*{-0.5mm} \log_2 \hspace*{-0.5mm} \Big( \hspace*{-0.5mm}1 \hspace*{-0.5mm} + \hspace*{-0.5mm} \frac{H_{m}^i p_{m}^i } {F_{r,m}^i q_r^i \hspace*{-0.5mm} + \hspace*{-0.5mm} 1}\Big)
\hspace*{-0.5mm} +\hspace*{-0.5mm} \mu_r \hspace*{-0.5mm} \log_2 \hspace*{-0.5mm} \Big(1 \hspace*{-0.5mm}+\hspace*{-0.5mm}  \frac{G_{r}^i q_{r}^i } {\rho L_{\mathrm{SI}}^i p_m^i \hspace*{-0.5mm}+ \hspace*{-0.5mm} 1}\Big)\hspace*{-0.5mm} \Bigg],\notag
\end{eqnarray}
where $H_{m}^i=\frac{\varpi_m\abs{h_m^i}^2}{\sigma_{\mathrm{z}_{\mathrm{DL}_m}}^2}$, $G_{r}^i=\frac{\varrho_r\abs{g_r^i}^2}{\sigma_{\mathrm{z}_{\mathrm{BS}}}^2}$, $F_{r,m}^i=\frac{\vartheta_{r,m}\abs{f_{r,m}^i}^2}{\sigma_{\mathrm{z}_{\mathrm{DL}_m}}^2}$, and $L_{\mathrm{SI}}^i=\frac{\abs{l_{\mathrm{SI}}^i}^2}{\sigma_{\mathrm{z}_{\mathrm{BS}}}^2}$, respectively.  $0\leq\rho\leq 1$  is a constant modelling the quality of the SI cancellation at the FD BS. $s_{m,r}^i\in\{0,1\}$ is the binary subcarrier allocation indicator. Specifically, $s_{m,r}^i=1$ if DL user $m$ and UL user $r$ are multiplexed on subcarrier $i$. Otherwise, $s_{m,r}^i=0$. Constants $0 \le w_m \le 1$ and $0 \le \mu_r \le 1$ are the resource allocation weight for DL user $m$ and UL user $r$, respectively.

\vspace*{-3mm}
\subsection{Optimization Problem Formulation}
The system objective is to maximize the weighted sum of the entire system throughput. The optimal joint power and subcarrier allocation policy is obtained by solving the following optimization problem:\vspace*{-2mm}
\begin{eqnarray} \label{pro}
&& \hspace*{-0mm} \underset{p_{m}^i,q_{r}^i,s_{m,r}^{i}}{\maxo} \sum_{i=1}^{N_{\mathrm{F}}}\sum_{m=1}^{K}  \sum_{r=1}^{J} U_{m,r}^i(p_{m}^i,q_{r}^i,s_{m,r}^{i})   \\
 \notag\mbox{s.t. }
&& \hspace*{-7mm} \mbox{C1: }\overset{N_{\mathrm{F}}}{\underset{i=1}{\sum}} \overset{K}{\underset{m=1}{\sum}}  \overset{J}{\underset{r=1}{\sum}} s_{m,r}^i p_{m}^i \le P_{\mathrm{max}}^{\mathrm{DL}}, \quad \mbox{C2: } p_m^i \ge 0, \,\, \forall i,m, \notag \\[-1mm]
&& \hspace*{-7mm} \mbox{C3: }\overset{N_{\mathrm{F}}}{\underset{i=1}{\sum}} \overset{K}{\underset{m=1}{\sum}}   s_{m,r}^i q_{r}^i \le P_{\mathrm{max}_r}^{\mathrm{UL}},\, \forall r, \quad \hspace*{-0mm} \mbox{C4: } q_r^i \ge 0, \,\, \forall i,r. \notag \\[-2mm]
&& \hspace*{-7mm} \mbox{C5: } s_{m,r}^i \in \{0,1\},\,\,\ \forall i,m,r, \quad \hspace*{-0mm} \mbox{C6: } \overset{K}{\underset{m=1}{\sum}}  \overset{J}{\underset{r=1}{\sum}} s_{m,r}^i \le 1, \forall i. \notag
\end{eqnarray}
Constraint C1 is the power constraint for the BS with a maximum transmit power allowance of $P_{\mathrm{max}}^{\mathrm{DL}}$.
Constraint C3 limits the transmit power of UL user $r$ to $ P_{\mathrm{max}_r}^{\mathrm{UL}}$.
Constraints C5 and C6 are imposed to guarantee that each subcarrier is allocated to at most one DL user and one UL user.
Here, we note that UL-to-DL user pairing is performed on every subcarrier.
Constraints C2 and C4 are the non-negative transmit power constraints for the DL and UL users, respectively.

The problem in \eqref{pro} is a mixed combinatorial non-convex problem due to the integer constraint for subcarrier allocation in C5 and the non-convex objective function. In general, there is no systematic approach for solving these problems efficiently. In some cases,
an exhaustive search or branch-and-bound
method is needed to obtain the globally optimal solution which
is computationally infeasible even for small $N_{\mathrm{F}}$, $K$, and $J$.  Therefore, in the next section, we propose a computational efficient suboptimal algorithm  based on successive convex approximation  which obtains a locally optimal solution for the optimization problem in \eqref{pro}.

\vspace*{-3mm}
\subsection{Joint Power and Subcarrier Allocation Algorithm}
To facilitate the presentation of the resource allocation algorithm design, in the sequel,  we rewrite the weighted throughput on subcarrier $i$ in \eqref{throughput_i} in an equivalent form:  \vspace*{-2mm}
\begin{eqnarray}
&&\hspace*{-6mm}\tilde{U}_{m,r}^i(\tilde{\mathbf{p}},\tilde{\mathbf{q}})\hspace*{-0.5mm} \notag\\[-1mm]
=&&\hspace*{-6mm}\hspace*{-0.5mm}w_m \hspace*{-0.5mm} \log_2\hspace*{-0.5mm} \Big(1 \hspace*{-1mm}+ \hspace*{-1mm} \frac{s_{m,r}^{i} H_{m}^i p_{m}^i } {s_{m,r}^{i}H_{m}^i p_{n}^i \hspace*{-1mm} +\hspace*{-1mm} 1}\Big) \hspace*{-1mm} +\hspace*{-0.5mm} \mu_r\log_2(1 \hspace*{-1mm}+ \hspace*{-1mm}s_{m,r}^{i}H_{n}^i q_{n}^i ) \notag\\
=&&\hspace*{-6mm}\hspace*{-0.5mm}w_m \hspace*{-0.5mm} \log_2\hspace*{-0.5mm} \Big(1 \hspace*{-1mm}+ \hspace*{-1mm} \frac{H_{m}^i \tilde{p}_{m,r}^i } {H_{m}^i \tilde{p}_{m,r}^i \hspace*{-1mm} +\hspace*{-1mm} 1}\Big) \hspace*{-1mm} +\hspace*{-0.5mm} \mu_r\log_2(1 \hspace*{-1mm}+ \hspace*{-1mm}H_{n}^i \tilde{q}_{m,r}^i )
\end{eqnarray}
where
$\tilde{p}_{m,r}^i=s_{m,r}^i p_{m}^i$ and $\tilde{q}_{m,r}^i=s_{m,r}^i q_{r}^i$ are auxiliary variables.
 Then, the original problem in \eqref{pro} can be rewritten as \vspace*{-4mm}
\begin{eqnarray} \label{eqv-pro}
\hspace*{-5mm}&&\hspace*{-0mm}\underset{\tilde{\mathbf{p}}, \tilde{\mathbf{q}}, \mathbf{s}}{\maxo}\,\, \,\, \notag \sum_{i=1}^{N_{\mathrm{F}}}\sum_{m=1}^{K} \sum_{r=1}^{J} \tilde{U}_{m,r}^i(\tilde{\mathbf{p}},\tilde{\mathbf{q}}) \notag \\[-1mm]
\hspace*{-5mm}&&\hspace*{-4mm}\mbox{s.t.}  \,\,\mbox{C1: } \overset{N_{\mathrm{F}}}{\underset{i=1}{\sum}} \overset{K}{\underset{m=1}{\sum}}  \overset{J}{\underset{r=1}{\sum}} \tilde{p}_{m,r}^i\hspace*{-0mm} \hspace*{-0mm}\le \hspace*{-0mm}P_{\mathrm{max}}^{\mathrm{DL}}, \, \mbox{C2: }\tilde{p}_{m,r}^i \ge 0, \,\, \forall m,r,i, \notag \\[-1mm]
&&\hspace*{1mm}\mbox{C3: } \overset{N_{\mathrm{F}}}{\underset{i=1}{\sum}} \overset{K}{\underset{m=1}{\sum}}  \tilde{q}_{m,r}^i\hspace*{-0mm} \hspace*{-0mm}\le \hspace*{-0mm}P_{\mathrm{max}}^{\mathrm{UL}}, \forall r, \,  \mbox{C4: }\tilde{q}_{m,r}^i \ge 0, \,\, \forall m,r,i, \notag \\[-1mm]
&&\hspace*{1mm}\mbox{C5, C6, }
\end{eqnarray}
where
$\tilde{\mathbf{p}}\in\mathbb{R}^{N_{\mathrm{F}}KJ \times1}$, $\tilde{\mathbf{q}}\in\mathbb{R}^{N_{\mathrm{F}}KJ \times1}$, and $\mathbf{s}\in\mathbb{Z}^{N_{\mathrm{F}}KJ \times1}$ are the collection of all $\tilde{p}_{m,r}^i$, $\tilde{q}_{m,r}^i$, and $s_{m,r}^{i}$, respectively.

We note that the product terms $\tilde{p}_{m,r}^i=s_{m,r}^i p_{m}^i$ and $\tilde{q}_{m,r}^i=s_{m,r}^i q_{r}^i$ in \eqref{eqv-pro} are the obstacles for the design of a computationally efficient resource allocation algorithm. In order to circumvent this difficulty, we adopt the big-M formulation to decompose the product terms \cite{lee2011mixed}. In particular, we impose the following additional constraints:
\begin{eqnarray}
&&\hspace*{-9mm}\mbox{C7: } \tilde{p}_{m,r}^i \le P_{\mathrm{max}}^{\mathrm{DL}} s_{m,r}^i, \,\, \forall i,m,r,\\
&&\hspace*{-9mm} \mbox{C8: } \tilde{p}_{m,r}^i \le p_m^i, \,\, \forall i,m,r,   \\
&&\hspace*{-9mm}\mbox{C9: } \tilde{p}_{m,r}^i \ge p_m^i  - (1  - s_{m,r}^i)P_{\mathrm{max}}^{\mathrm{DL}} ,\,\, \forall i,m,r, \\
&&\hspace*{-9mm} \mbox{C10: }  \tilde{p}_{m,r}^i \ge 0, \forall i,m,r, \\
&&\hspace*{-9mm} \mbox{C11: } \tilde{q}_{m,r}^i \le P_{\mathrm{max}_r}^{\mathrm{UL}} s_{m,r}^i,  \forall i,m,r, \\
&&\hspace*{-9mm}\mbox{C12: } \tilde{q}_{m,r}^i \le q_r^i, \,\, \forall i,m,r,   \quad \mbox{C13: } \tilde{q}_{m,r}^i \ge 0, \,\, \forall i,m,r, \\
&&\hspace*{-9mm}\mbox{C14: } \tilde{q}_{m,r}^i \ge q_r^i  - (1  - s_{m,r}^i)P_{\mathrm{max}_r}^{\mathrm{UL}} ,\,\, \forall i,m,r.
\end{eqnarray}
Besides,  constraint C5 in  \eqref{eqv-pro} is a combinatorial constraint. Hence, we rewrite constraint C5 in the following equivalent form: \vspace*{-3mm}
\begin{eqnarray}
&&\hspace*{-10mm}\text{C5}\mbox{a: } \overset{{N_{\mathrm{F}}}}{\underset{i=1}{\sum}} \overset{K}{\underset{m=1}{\sum}}  \overset{J}{\underset{r=1}{\sum}} s_{m,r}^i - (s_{m,r}^i)^2 \le 0 \,\,\,\, \text{and} \\
&&\hspace*{-10mm}\text{C5}\mbox{b: } 0 \le s_{m,r}^i \le 1,\,\, \forall i,m,r.
\end{eqnarray}
Now,  $s_{m,r}^i$ is a continuous optimization variable having values between zero and one.
We note that constraint $\mbox{C5a}$ is the difference of two convex functions which is known as a reverse convex function \cite{che2014Joint}\nocite{ng2015power}--\cite{dinh2010local}. In order to handle constraint $\mbox{C5a}$, we reformulate the problem in \eqref{eqv-pro} as \vspace*{-2mm}
\begin{eqnarray} \label{penalty-pro}
\hspace*{-5mm}&&\hspace*{-5mm}\underset{\tilde{\mathbf{p}},\tilde{\mathbf{q}},\mathbf{s}}{\mino}\,\, \,\, \notag \sum_{i=1}^{{N_{\mathrm{F}}}}\sum_{m=1}^{K}  \overset{J}{\underset{r=1}{\sum}}  -U_{m,r}^i(\tilde{\mathbf{p}},\tilde{\mathbf{q}})\notag \\
&&\hspace*{11mm}+ \eta\Big(\overset{{N_{\mathrm{F}}}}{\underset{i=1}{\sum}} \overset{K}{\underset{m=1}{\sum}} \overset{K}{\underset{n=1}{\sum}}s_{m,r}^i -\overset{{N_{\mathrm{F}}}}{\underset{i=1}{\sum}} \overset{K}{\underset{m=1}{\sum}} \overset{K}{\underset{n=1}{\sum}}(s_{m,r}^i)^2\Big) \notag \\
\hspace*{-1mm}&&\hspace*{4mm}\mbox{s.t.}  \hspace*{7mm}\,\,\mbox{C1--C4}, \text{C5}\mbox{b}, \mbox{C6--C14},
\end{eqnarray}
where $\eta \gg 1$ is a large constant which acts as a penalty factor to penalize the objective function for any $s_{m,n}^i$ that is not equal to zero or one. It can be shown that the problems in \eqref{penalty-pro} and \eqref{eqv-pro} are equivalent for $\eta \gg 1$ \cite{che2014Joint,ng2015power}. Therefore, we consider the resource allocation algorithm design for \eqref{penalty-pro}.

\begin{table}
 \vspace*{-2mm}
\begin{algorithm} [H]                  
\caption{Successive Convex Approximation}          
\label{alg1}                           
\begin{algorithmic} [1]
\small          
\STATE Initialize the maximum number of iterations $I_{\mathrm{max}}$, penalty factor $\eta \gg 1$, iteration index $k=1$, and initial point $\tilde{\mathbf{p}}^{(1)}$, $\tilde{\mathbf{q}}^{(1)}$, and $\mathbf{s}^{(1)}$

\REPEAT
\STATE Solve \eqref{dc} for a given $\tilde{\mathbf{p}}^{(k)}$, $\tilde{\mathbf{q}}^{(k)}$, and $\mathbf{s}^{(k)}$ and store the intermediate resource allocation policy $\{\tilde{\mathbf{p}}, \tilde{\mathbf{q}}, \mathbf{s}\}$

\STATE Set $k=k+1$ and $\tilde{\mathbf{p}}^{(k)}=\tilde{\mathbf{p}}$, $\tilde{\mathbf{q}}^{(k)}=\tilde{\mathbf{q}}$, and $\mathbf{s}^{(k)}=\mathbf{s}$

\UNTIL convergence or $k=I_{\mathrm{max}}$

\STATE $\tilde{\mathbf{p}}^{*}=\tilde{\mathbf{p}}^{(k)}$, $\tilde{\mathbf{q}}^{*}=\tilde{\mathbf{q}}^{(k)}$, and $\mathbf{s}^{*}=\mathbf{s}^{(k)}$

\end{algorithmic}
\end{algorithm}\vspace*{-7mm}
\end{table}

Although the constraints in \eqref{penalty-pro} span a convex set, the optimization problem in \eqref{penalty-pro} is still non-convex because of the objective function. To handle this difficulty, we rewrite the problem as \vspace*{-0mm}
\begin{eqnarray}\label{dc-penalty-pro}
\hspace*{-1mm}&&\hspace*{-0mm}\underset{\tilde{\mathbf{p}},\tilde{\mathbf{q}},\mathbf{s}}{\mino}\,\, \,\, F(\tilde{\mathbf{p}},\tilde{\mathbf{q}})-G(\tilde{\mathbf{p}},\tilde{\mathbf{q}})+\eta(H(\mathbf{s})-M(\mathbf{s})) \notag \\
\hspace*{-1mm}&&\hspace*{4mm}\mbox{s.t.}  \hspace*{7mm}\,\,\mbox{C1--C3}, \text{C4}\mbox{b}, \mbox{C5--C15},
\end{eqnarray}
where \vspace*{-2mm}
\begin{eqnarray}
\hspace*{-0mm}F(\tilde{\mathbf{p}},\tilde{\mathbf{q}})
\hspace*{-3mm}&=&\hspace*{-3mm}\sum_{i=1}^{{N_{\mathrm{F}}}}\sum_{m=1}^{K}  \overset{J}{\underset{r=1}{\sum}} \hspace*{-1mm} - \hspace*{-1mm} w_m \hspace*{-0.5mm} \log_2\hspace*{-1mm} \big(1 \hspace*{-1mm}+ \hspace*{-1mm} H_{m}^i \tilde{p}_{m,r}^i \hspace*{-1mm} + \hspace*{-1mm} F_{r,m}^i \tilde{q}_{m,r}^i   \big)  \hspace*{-1mm} \notag \\
\hspace*{-5mm}&-&\hspace*{-3mm} \mu_r\log_2\big(1 \hspace*{-0mm}+ \hspace*{-0mm} G_{r}^i \tilde{q}_{m,r}^i + \rho L_{\mathrm{SI}}^i\tilde{p}_{m,r}^i \big)\\
\hspace*{-3mm} H(\mathbf{s}) \hspace*{-3mm}&=&\hspace*{-3mm} \overset{{N_{\mathrm{F}}}}{\underset{i=1}{\sum}} \overset{K}{\underset{m=1}{\sum}} \overset{J}{\underset{r=1}{\sum}} s_{m,r}^i, \, M(\mathbf{s}) \hspace*{-1mm}=\hspace*{-1mm}\overset{{N_{\mathrm{F}}}}{\underset{i=1}{\sum}} \overset{K}{\underset{m=1}{\sum}} \overset{J}{\underset{r=1}{\sum}}  (s_{m,r}^i)^2, \, \,\,\,\,\,\,\,\ \\
\hspace*{-2mm}G(\tilde{\mathbf{p}},\tilde{\mathbf{q}})\hspace*{-3mm}
&=&\hspace*{-3mm}\sum_{i=1}^{{N_{\mathrm{F}}}}\sum_{m=1}^{K} \overset{J}{\underset{r=1}{\sum}}  -w_m \hspace*{-0.5mm} \log_2\hspace*{0mm} (1 + F_{r,m}^i \tilde{q}_{m,r}^i )\notag \\
\hspace*{-5mm}&-&\hspace*{-3mm} \mu_r \hspace*{-0.5mm} \log_2\hspace*{0mm} \big(1 + \rho L_{\mathrm{SI}}^i \tilde{p}_{m,r}^i \big).
\end{eqnarray}\vspace*{-4mm}

We note that $F(\tilde{\mathbf{p}},\tilde{\mathbf{q}})$, $G(\tilde{\mathbf{p}},\tilde{\mathbf{q}})$, $H(\mathbf{s})$, and $M(\mathbf{s})$ are convex functions and the problem in \eqref{dc-penalty-pro} belongs to the class of difference of convex (d.c.) function programming. As a result, we can apply successive convex approximation \cite{dinh2010local} to obtain a locally optimal solution for \eqref{dc-penalty-pro}.
Since $G(\tilde{\mathbf{p}},\tilde{\mathbf{q}})$ and $M(\mathbf{s})$ are differentiable convex functions, for any feasible point $\tilde{\mathbf{p}}^{(k)}$, $\tilde{\mathbf{q}}^{(k)}$, and $\mathbf{s}^{(k)}$, we have the following inequalities: \vspace*{-4mm}
\begin{eqnarray}\label{ineq1}
G(\tilde{\mathbf{p}},\tilde{\mathbf{q}}) \hspace*{-3mm}&\ge&\hspace*{-3mm} G(\tilde{\mathbf{p}}^{(k)},\tilde{\mathbf{q}}^{(k)}) \hspace*{-0.7mm}+\hspace*{-0.7mm}\nabla_{\tilde{\mathbf{p}}} G(\tilde{\mathbf{p}}^{(k)},\tilde{\mathbf{q}}^{(k)})^T(\tilde{\mathbf{p}}-\tilde{\mathbf{p}}^{(k)})\notag \\
\hspace*{-3mm}&+&\hspace*{-3mm} \nabla_{\tilde{\mathbf{q}}} G(\tilde{\mathbf{p}}^{(k)},\tilde{\mathbf{q}}^{(k)})^T(\tilde{\mathbf{q}}-\tilde{\mathbf{q}}^{(k)})\,\,\, \text{and}\\
\label{ineq2}M(\mathbf{s}) \hspace*{-3mm}&\ge&\hspace*{-3mm} M(\mathbf{s}^{(k)}) +\nabla_{\mathbf{s}} M(\mathbf{s}^{(k)})^T(\mathbf{s}-\mathbf{s}^{(k)}),
\end{eqnarray}
where the right hand sides of \eqref{ineq1} and \eqref{ineq2} are affine functions representing the global underestimation of $G(\tilde{\mathbf{p}},\tilde{\mathbf{q}})$ and $M(\mathbf{s})$, respectively.
Therefore, for any given $\tilde{\mathbf{p}}^{(k)}$, $\tilde{\mathbf{q}}^{(k)}$, and $\mathbf{s}^{(k)}$, we can obtain an upper bound for \eqref{dc-penalty-pro} by solving the following convex optimization problem:
\begin{eqnarray}\label{dc}
\hspace*{-7mm}&&\hspace*{-1mm} \underset{\tilde{\mathbf{p}},\tilde{\mathbf{q}},\mathbf{s}}{\mino} F(\tilde{\mathbf{p}},\hspace*{-0.7mm}  \tilde{\mathbf{q}}) \hspace*{-0.7mm} - \hspace*{-0.7mm} G(\tilde{\mathbf{p}}^{(k)}\hspace*{-0.7mm} ,\hspace*{-0.7mm} \tilde{\mathbf{q}}^{(k)}) \hspace*{-0.7mm} -\hspace*{-0.7mm} \nabla_{\tilde{\mathbf{p}}} G(\tilde{\mathbf{p}}^{(k)}\hspace*{-0.7mm} ,\hspace*{-0.7mm} \tilde{\mathbf{q}}^{(k)})^T(\tilde{\mathbf{p}}\hspace*{-0.7mm} -\hspace*{-0.7mm} \tilde{\mathbf{p}}^{(k)})  \notag \\
\hspace*{-7mm}&&\hspace*{11mm}-\hspace*{-0.7mm} \nabla_{\tilde{\mathbf{q}}} G(\tilde{\mathbf{p}}^{(k)},\tilde{\mathbf{q}}^{(k)})^T(\tilde{\mathbf{q}}-\tilde{\mathbf{q}}^{(k)})
\hspace*{-0.7mm}+\hspace*{-0.7mm}\eta\big(H(\mathbf{s})-M(\mathbf{s}^{(k)}) \notag \\
 \hspace*{-7mm}&&\hspace*{11mm}-\hspace*{-0.7mm} \nabla_{\mathbf{s}} M(\mathbf{s}^{(k)})^T(\mathbf{s}-\mathbf{s}^{(k)})\big) \notag \\
\hspace*{-7mm}&&\hspace*{4mm}\mbox{s.t.}  \hspace*{7mm}\,\,\mbox{C1--C4}, \text{C5}\mbox{b}, \mbox{C6--C14},
\end{eqnarray}
where \vspace*{-3mm}
\begin{eqnarray}
\hspace*{-13mm}&&\hspace*{-0mm} \nabla_{\tilde{\mathbf{p}}}  G(\tilde{\mathbf{p}}^{(k)},\tilde{\mathbf{q}}^{(k)} )^T (\tilde{\mathbf{p}} - \tilde{\mathbf{p}}^{(k)} ) \notag \\
\hspace*{-13mm}&=&\hspace*{-0mm}\overset{{N_{\mathrm{F}}}}{\underset{i=1}{\sum}} \overset{K}{\underset{m=1}{\sum}} \overset{J}{\underset{r=1}{\sum}}  -\hspace*{-0mm} \frac{\mu_r \rho L_{\mathrm{SI}}^i(\tilde{p}_{m,r}^i-\tilde{p}_{m,r}^{i(k)})}{\big(1+ \rho L_{\mathrm{SI}}^i\tilde{p}_{m,r}^{i(k)}\big)\ln(2)}, \\
\hspace*{-13mm}&&\hspace*{-0mm} \nabla_{\tilde{\mathbf{q}}}  G( \tilde{\mathbf{p}}^{(k)},\tilde{\mathbf{q}}^{(k)})^T  (\tilde{\mathbf{q}} - \tilde{\mathbf{q}}^{(k)} ) \notag \\
\hspace*{-13mm}&=&\hspace*{-0mm}\overset{{N_{\mathrm{F}}}}{\underset{i=1}{\sum}} \overset{K}{\underset{m=1}{\sum}} \overset{J}{\underset{r=1}{\sum}}
- \frac{w_m  F_{r,m}^i (\tilde{q}_{m,r}^i - \tilde{q}_{m,r}^{i(k)})}{(1+ F_{r,m}^i \tilde{q}_{m,r}^{i(k)}   )\ln(2)} \\
\hspace*{-20mm} && \nabla_{{\mathbf{s}}} M( \mathbf{s}^{(k)})^T  (\mathbf{s} - \mathbf{s}^{(k)} )\notag \\
\hspace*{-20mm}&=& \overset{{N_{\mathrm{F}}}}{\underset{i=1}{\sum}} \overset{K}{\underset{m=1}{\sum}} \overset{J}{\underset{r=1}{\sum}} 2s_{m,r}^{i (k)}(s_{m,r}^{i}-s_{m,r}^{i (k)}).
\end{eqnarray}
Then, we employ an iterative algorithm to tighten the obtained upper bound as summarized in \textbf{Algorithm 1}.
In each iteration, the convex problem in \eqref{dc} can be solved efficiently by standard convex program solvers such as CVX \cite{website:CVX}.
By solving the convex upper bound problem in \eqref{dc}, the proposed iterative scheme generates a sequence of feasible solutions $\tilde{\mathbf{p}}^{(k+1)}$, $\tilde{\mathbf{q}}^{(k+1)}$, and $\mathbf{s}^{(k+1)}$ successively. The proposed suboptimal iterative algorithm converges to a locally optimal solution of \eqref{dc} with  polynomial time computational complexity \cite{dinh2010local}.

\vspace*{-5mm}
\section{Simulation Results}
\begin{table}[t]\vspace*{-0mm}\caption{Simulation Parameters.}\vspace*{-0mm}\label{tab:parameters} 
\newcommand{\tabincell}[2]{\begin{tabular}{@{}#1@{}}#2\end{tabular}} \small \small
\centering
\begin{tabular}{|l|l|}\hline
\hspace*{-1mm}Carrier center frequency& $2.5$ GHz  \\
\hline
\hspace*{-1mm}System bandwidth & $5$ MHz \\
\hline
\hspace*{-1mm}Number of subcarriers, ${N_{\mathrm{F}}}$ & $64$  \\
\hline
\hspace*{-1mm}Bandwidth per subcarrier & $78$ kHz \\
\hline
\hspace*{-1mm}Path loss exponent  &  \mbox{$3.6$}    \\
\hline
\hspace*{-1mm}SI cancellation constant, $\rho$ &    \mbox{$-90$ dB}   \\
\hline
\hspace*{-1mm}DL user noise power, $\sigma_{\mathrm{z}_{{\mathrm{DL}_m}}}^2$  &  \mbox{$-125$ dBm}   \\
\hline
\hspace*{-1mm}UL BS noise power, $\sigma_{\mathrm{z}_{{\mathrm{BS}}}}^2$ &  \mbox{$-125$ dBm}   \\
\hline
\hspace*{-1mm}Maximum transmit power for UL users, $P_{\mathrm{max}_r}^{\mathrm{UL}}$ & \mbox{$18$ dBm}   \\
\hline
\hspace*{-1mm} BS antenna gain &  \mbox{$10$ dBi}   \\
\hline
\end{tabular}
\vspace*{-1mm}
\end{table}
\vspace*{-1mm}

In this section, we investigate the performance of the proposed resource allocation scheme through simulations.
We adopt the simulation parameters given in Table \ref{tab:parameters}, unless specified otherwise.
A single cell with two ring-shaped boundary regions is considered. The outer boundary and the inner boundary have radii of $30$ meters and $600$ meters, respectively. The $K$ DL and $J$ UL users are randomly and uniformly distributed between the inner and the outer boundary.
The FD BS is located at the center of the cell.
The maximum transmit power of the FD BS is $P_{\mathrm{max}}^{\mathrm{DL}}$.
We set the same weight for all the users, i.e., $w_m=\mu_r=1,\forall m, r$.
The penalty term $\eta$ for the proposed algorithm is set to $10 \log_2(1+\frac{P_{\mathrm{max}}^{\mathrm{DL}}}{\sigma_{\mathrm{z}_{{\mathrm{DL}}_m}}^2})$.
The small-scale fading of the DL channels, the UL channels, and the channel between the DL and UL users is modeled as independent and identically distributed Rayleigh fading. The fading coefficients of the SI channel on each subcarrier are generated as independent and identically distributed Rician random variables with Rician factor $5$ dB. The results shown in this section were averaged over different realizations of both path loss and multipath fading.

\begin{figure}[t]
 \centering\vspace*{-3mm}
\includegraphics[width=3.4in]{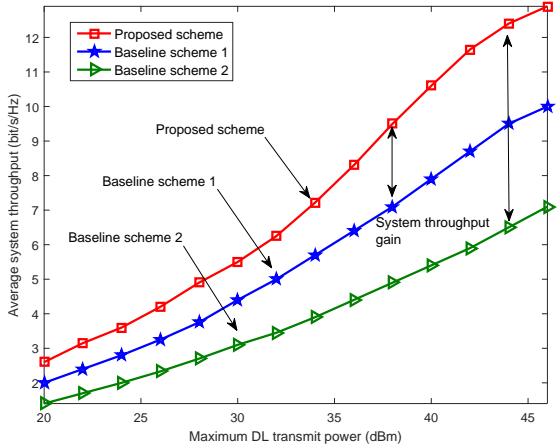} \vspace*{-8mm}
\caption{Average system throughput (bits/s/Hz) versus the maximum DL transmit power at the BS (dBm), $P_{\mathrm{max}}^{\mathrm{DL}}$, for different resource allocation schemes and $K=J=10$. The double-sided arrows indicate the performance gain brought by the proposed scheme.} \label{fig:wsr_vs_power}\vspace*{-5mm}
\end{figure}

In Figure \ref{fig:wsr_vs_power}, we investigate the average system throughput versus  the maximum transmit power at the FD BS, $P_{\mathrm{max}}^{\mathrm{DL}}$, for $K=10$ DL users and $J=10$ UL users. The number of iterations for the proposed
iterative resource allocation algorithm is
$5$. As can be observed, the average system throughput increases monotonically with the maximum transmit power $P_{\mathrm{max}}^{\mathrm{DL}}$. In fact, the proposed scheme is able to obtain a locally optimal solution of \eqref{pro} which can effectively exploit the increased transmit power budget to improve the received signal-to-interference-plus-noise ration (SINR) at the users. For comparison, Figure \ref{fig:wsr_vs_power} also shows the average system throughput of two baseline schemes.
For baseline scheme $1$, we adopt the suboptimal power and subcarrier allocation for the considered FD MC system proposed in \cite{Nam15Joint}.
For baseline scheme $2$, a traditional HD MC system is considered.
As can be observed, the proposed scheme achieves a considerably higher average system throughput than baseline scheme $1$ due to the joint power and subcarrier allocation. Besides, baseline scheme $2$ can only achieve a substantially lower average system throughput compared to the proposed scheme due to its underutilization of the spectral resource. For instance, for $P_{\mathrm{max}}^{\mathrm{DL}}=46$ dBm, the proposed scheme achieves roughly a $30\%$ and $86\%$ higher average system throughput than baseline schemes $1$ and $2$, respectively.

\begin{figure}[t]
 \centering\vspace*{-3mm}
\includegraphics[width=3.4in]{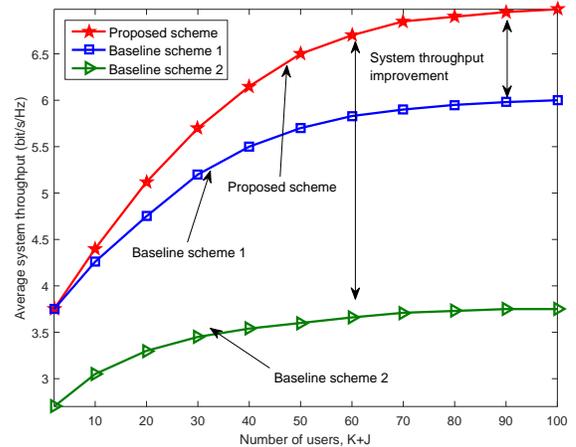} \vspace*{-8mm}
\caption{Average system throughput (bits/s/Hz) versus  the number of users for different resource allocation schemes with $P_{\mathrm{max}}^{\mathrm{DL}}=45$ dBm. }\label{fig:wsr_vs_usernum}\vspace*{-5mm}
\end{figure}
In Figure \ref{fig:wsr_vs_usernum}, we investigate the average system throughput versus  the number of users for a maximum transmit power of $P_{\mathrm{max}}^{\mathrm{DL}}=31$ dBm. We assume that there are equal numbers of DL and UL users in the system. As can be observed, the average system throughput for the proposed scheme and baseline schemes increase with the number of users since all schemes are able to exploit multiuser diversity. Besides, it can be observed from Figure \ref{fig:wsr_vs_usernum} that the average system throughput of the proposed scheme grows faster with an increasing number of users than that of the baseline schemes.
In fact, baseline scheme 1 divides the subcarrier and power allocation problem into two individual sub-problems which leads to performance degradation. In particular, the performance gap between the proposed scheme and baseline scheme 1 is enlarged for an increasing number of users. On the other hand, baseline scheme $2$ achieves a lower average system throughput compared to the proposed scheme and baseline 1, since DL and UL transmission are separated into orthogonal radio resources which leads to inefficient spectrum utilization.

\vspace*{-2mm}
\section{Conclusion}
In this paper, we studied the joint power and subcarrier allocation for an FD MC system. The resource allocation design was formulated as a non-convex optimization problem with the objective to maximize the weighted system throughput. A suboptimal resource allocation algorithm design based on successive convex approximation was proposed. Simulation results unveiled that the proposed scheme for FD MC systems achieves a significant improvement in system performance compared to two baseline schemes.

\vfill\pagebreak


\begin{thebibliography}{10}

\bibitem{book:david_wirelss_com}
D.~Tse and P.~Viswanath,
\newblock {\em {Fundamentals of Wireless Communication}},
\newblock {Cambridge University Pres}, first edition, 2005.

\bibitem{JR:Linglong_OFDM}
L.~Dai, Z.~Wang, and Z.~Yang,
\newblock ``{Time-Frequency Training OFDM with High Spectral Efficiency and
  Reliable Performance in High Speed Environments},''
\newblock vol. 30, no. 4, pp. 695--707, May 2012.

\bibitem{JR:Linglong_OFDM2}
L.~Dai, J.~Wang, Z.~Wang, P.~Tsiaflakis, and M.~Moonen,
\newblock ``{Spectrum- and Energy-Efficient OFDM Based on Simultaneous
  Multi-Channel Reconstruction},''
\newblock {\em IEEE Trans. Signal Process.}, vol. 61, no. 23, pp. 6047--6059,
  Dec 2013.

\bibitem{JR:Chen_MC}
X.~Chen and X.~Wang,
\newblock ``{Statistical Precoder Design for Space-Time-Frequency Block Codes
  in Multiuser MISO-MC-CDMA Systems},''
\newblock {\em IEEE Systems Journal}, vol. 10, no. 1, pp. 218--227, Mar. 2016.

\bibitem{JR:Roger_OFDMA}
C.~Y. Wong, R.~S. Cheng, K.~B. Letaief, and R.~D. Murch,
\newblock ``{Multiuser OFDM with Adaptive Subcarrier, Bit, and Power
  Allocation},''
\newblock {\em IEEE J. Select. Areas Commun.}, vol. 17, pp. 1747--1758, Oct.
  1999.

\bibitem{JR:QQ_OFDM_EE}
Q.~Wu, M.~Tao, and W.~Chen,
\newblock ``{Joint Tx/Rx Energy-Efficient Scheduling in Multi-Radio Wireless
  Networks: A Divide-and-Conquer Approach},''
\newblock {\em IEEE Trans. Wireless Commun.}, vol. 15, no. 4, pp. 2727--2740,
  Apr. 2016.

\bibitem{ng2012energy}
D.~W.~K. Ng, E.~S. Lo, and R.~Schober,
\newblock ``{Energy-Efficient Resource Allocation in Multi-Cell OFDMA Systems
  with Limited Backhaul Capacity},''
\newblock {\em IEEE Trans. Wireless Commun.}, vol. 11, no. 10, pp. 3618--3631,
  Sept. 2012.

\bibitem{sun2016optimal}
Y.~Sun, D.~W.~K. Ng, Z.~Ding, and R.~Schober,
\newblock ``{Optimal Joint Power and Subcarrier Allocation for MC-NOMA
  Systems},'' {[Online] }\url{http://arxiv.org/abs/1503.06021},
\newblock accepted for presentation at IEEE Global Commun. Conf. 2016.

\bibitem{CN:MC_NOMA_Zhiqiang}
Z.~Wei, D.~W.~K. Ng, and J.~Yuan,
\newblock ``{Power-Efficient Resource Allocation for {MC-NOMA} with Statistical
  Channel State Information},'' {[Online]
  }\url{http://arxiv.org/abs/1607.01116}, 2016,
\newblock accepted for presentation at IEEE Global Commun. Conf. 2016.

\bibitem{Sun16FDSecurity}
Y.~Sun, D.~W.~K. Ng, J.~Zhu, and R.~Schober,
\newblock ``{Multi-Objective Optimization for Robust Power Efficient and Secure
  Full-Duplex Wireless Communication Systems},''
\newblock {\em IEEE Trans. Wireless Commun.}, vol. 15, no. 8, pp. 5511--5526,
  2016.

\bibitem{Kwan2016FDDAS}
D.~W.~K. Ng, Y.~Wu, and R.~Schober,
\newblock ``{Power Efficient Resource Allocation for Full-Duplex Radio
  Distributed Antenna Networks},''
\newblock {\em IEEE Trans. Wireless Commun.}, vol. 15, no. 4, pp. 2896--2911,
  Apr. 2016.

\bibitem{Dynamic_FDrelay}
D.~W.~K. Ng, E.~S. Lo, and R.~Schober,
\newblock ``{Dynamic Resource Allocation in MIMO-OFDMA Systems with Full-Duplex
  and Hybrid Relaying},''
\newblock {\em IEEE Trans. Commun.}, vol. 60, no. 5, pp. 1291--1304, May 2012.

\bibitem{Nam15Joint}
C.~Nam, C.~Joo, and S.~Bahk,
\newblock ``{Joint Subcarrier Assignment and Power Allocation in Full-Duplex
  OFDMA Networks},''
\newblock {\em IEEE Trans. Wireless Commun.}, vol. 14, no. 6, pp. 3108--3119,
  Jun. 2015.

\bibitem{lee2011mixed}
J.~Lee and S.~Leyffer,
\newblock {\em {Mixed Integer Nonlinear Programming}},
\newblock Springer Science \& Business Media, 2011.

\bibitem{che2014Joint}
E.~Che, H.~D. Tuan, and H.~H. Nguyen,
\newblock ``{Joint Optimization of Cooperative Beamforming and Relay Assignment
  in Multi-User Wireless Relay Networks},''
\newblock {\em IEEE Trans. Wireless Commun.}, vol. 13, no. 10, pp. 5481--5495,
  Oct. 2014.

\bibitem{ng2015power}
D.~W.~K. Ng, Y.~Wu, and R.~Schober,
\newblock ``{Power Efficient Resource Allocation for Full-Duplex Radio
  Distributed Antenna Networks},''
\newblock {\em IEEE Trans. Wireless Commun.}, vol. 15, no. 4, pp. 2896--2911,
  Apr. 2016.

\bibitem{dinh2010local}
Q.~T. Dinh and M.~Diehl,
\newblock ``{Local Convergence of Sequential Convex Programming for Nonconvex
  Optimization},''
\newblock in {\em Recent Advances in Optimization and its Applications in
  Engineering}, pp. 93--102. Springer, 2010.

\bibitem{website:CVX}
M.~Grant and S.~Boyd,
\newblock ``{CVX: Matlab Software for Disciplined Convex Programming, version
  2.1},'' {[Online] }\url{http://cvxr.com/cvx}, Mar. 2014.

\end{thebibliography}
\end{document}